\documentclass[twocolumn,english,prl]{revtex4-1}
\usepackage[colorlinks=true,urlcolor=blue,citecolor=blue,linkcolor=blue]{hyperref} 
\usepackage[T1]{fontenc}
\usepackage[latin9]{inputenc}
\usepackage{amssymb}
\usepackage{graphicx}
\usepackage{amsmath,color}
\usepackage{mathrsfs}
\usepackage{float}
\usepackage{indentfirst}
\usepackage{braket}
\usepackage{txfonts}

\makeatletter

\def\journal #1, #2, #3, 1#4#5#6{{\sl #1~}{\bf #2}, #3 (1#4#5#6) }

\def\eqa{\begin{eqnarray}}
\def\eea{\end{eqnarray}}
\newcommand{\eq}{\begin{equation}}
\newcommand{\ee}{\end{equation}}

\newcommand{\Eq}[1]{Eq.~(\ref{#1})}

\makeatother

\usepackage{babel}

\begin{document}

%\title{Three-way Restricted Boltzmann Machines Can Learn Cluster Updates}
%\title{Discovering Cluster Updates with Boltzmann Machines}
\title{Can Boltzmann Machines Discover Cluster Updates ? }

\author{Lei Wang}
\email{wanglei@iphy.ac.cn}
\affiliation{Beijing National Lab for Condensed Matter Physics and Institute
of Physics, Chinese Academy of Sciences, Beijing 100190, China }

\begin{abstract}
Boltzmann machines are physics informed generative models with wide applications in machine learning. They can learn the probability distribution from an input dataset and generate new samples accordingly. Applying them back to physics, the Boltzmann machines are ideal recommender systems to accelerate Monte Carlo simulation of physical systems due to their flexibility and effectiveness. More intriguingly, we show that the generative sampling of the Boltzmann Machines can even discover unknown cluster Monte Carlo algorithms. The creative power comes from the latent representation of the Boltzmann machines, which learn to mediate complex interactions and identify clusters of the physical system. We demonstrate these findings with concrete examples of the classical Ising model with and without four spin plaquette interactions. Our results endorse a fresh research paradigm where intelligent machines are designed to create or inspire human discovery of innovative algorithms.  
%We design a restricted Boltzmann machine which can automatically learn cluster Monte Carlo updates for statistical physics problems.
%We construct an RBM by placing the hidden units on the links of the lattice and allow them interact with the two spins on the two ends of the bond. Recommender updates from the 3-way RBM automatically reproduce the cluster Monte Carlo updates. The sampled on and off these hidden units corresponds to ``freeze'' and ``delete'' operations. While for a given set of the hidden units, one can flip the spins of connected components. Training such an RBM to Monte Carlo data can speed Monte Carlo simulations. 
%BM offers a general framework for cluster Monte Carlo algorithms for statistical physics problems. 
%Conceptually, our work offers the hope that the machine learning approach can help us invent new algorithm one day.  
\end{abstract}
\maketitle

\paragraph{Introduction--} It is intriguing to wonder whether artificial intelligence can make scientific discoveries~\cite{langley1987scientific, alai2004ai, nilsson2009quest, gil2014amplify} ever since the dawn of AI. With recent rapid progress in machine learning, AI is reaching human-level performance in many tasks and it is becoming even more optimistic that AI can indeed achieve the lofty goal of making scientific discoveries. Focussing on the computational study of many-body physical problems with relatively well-defined rules and targets, e.g. distinguishing phases of matter or finding the lowest energy state, the above question is addressed recently in Refs.~\cite{Carrasquilla:2016wu, Nieuwenburg,  Wang:2016fub, Carleo:2016vp}. 

An equally interesting question is whether AI can invent, or at least, inspire human discovery of new problem-solving strategies, i.e. new algorithms. In this respect, the two examples from DeepMind, where computers discover optimal strategy in the video game Breakout~\cite{Mnih:2015jpb} and master the board game Go~\cite{Silver:2016hl}, are particularly inspiring. These are vivid examples of an intelligence agent which discovers nontrivial problem-solving strategies even unknown to its programmer. %or inventing new efficient algorithms to tackle challenging scientific problems. 
%caught heated attentions in recent years since they 
%This is inspiring because a key observations is that the programmer of the program may not be a good player of this particular game. In the end, by observing how the agent play, human learns better strategy of playing the game. %This requires we do not use the machine learning models as a black box manner but really can be comprehended by human and inspire. 
%and was demonstrated recently using the Newton's law as an example~\cite{Schmidt:2009dt}. 
%Next, one may wonder whether we can let intelligent machines invent new algorithms for us. %It is hard to imagine that all the algorithm has been invented.

Along this line, it is highly desirable to devise new efficient Monte Carlo algorithms to sample configuration spaces of physical problems more rapidly. The existing ones such as hybrid Monte Carlo~\cite{Duane:1987uq},  cluster algorithms~\cite{1987PhRvL..58...86S, Wolff:1989iy}, loop algorithm~\cite{PhysRevLett.70.875}, and worm algorithm~\cite{PhysRevLett.87.160601} are all landmark achievements in computational physics and find wide applications in physical, statistical and biological problems. There are some recent efforts to improve the Monte Carlo sampling~\cite{Huang:2017fg, Liu:2017da, 1611.09364, Huang:2016wx, 1612.03804} using ideas and techniques from machine learning. Similar approaches were also discussed in statistics literature~\cite{neal2012bayesian, liu2008monte, rasmussen2003gaussian} where one uses surrogate functions to guide and accelerate hybrid Monte Carlo calculations~\cite{Duane:1987uq}. As suggested in~\cite{Huang:2017fg}, using intelligent Boltzmann Machines (BM)~\cite{ackley1985learning, Hinton:1986ub} opens possibilities of algorithmic innovations because they can discovery unknown algorithmic strategies instead of merely acting as cheaper surrogate functions. 

\begin{figure}[!t]
\centering
\includegraphics[width=\columnwidth]{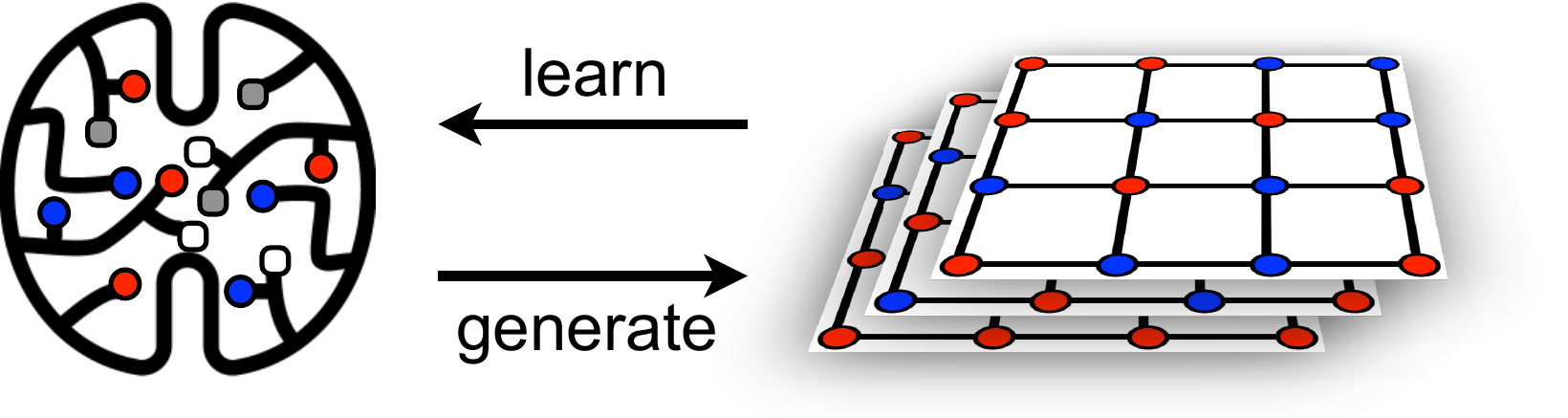}
\caption{A schematic plot of the Boltzmann Machine which can learn from data and generate new samples. The Boltzmann Machine consists of stochastic visible (red and blue dots) and hidden units (white and gray squares) connected into a network. The colors indicate the status of various units which follow a Boltzmann distribution with a given energy function. The BM learns marginal probability distribution of visible variables from data by tuning its energy function parameters. We show that BM with an appropriated designed architecture can discover efficient cluster Monte Carlo algorithms in the generative sampling.}
\label{fig:concept}
\end{figure}

%BM 
BM is an energy based model consists of stochastic visible ($\mathbf{s}$) and hidden ($\mathbf{h}$) variables illustrated in Fig~\ref{fig:concept}. The BM architecture is specified by an energy function $E(\mathbf{s}, \mathbf{h})$ which depends on the connectivity, connection weights, and biases of the units. The joint probability distribution of the units follows the Boltzmann distribution $p(\mathbf{s}, \mathbf{h})= e^{-E(\mathbf{s}, \mathbf{h})}$. One can train the BM by tuning its energy function such that the marginal distribution of the visible variables $p(\mathbf{s}) = \sum_{\mathbf{h}} p(\mathbf{s},\mathbf{h})$ approximates the target probability distribution $\pi(\mathbf{s})$ of a dataset. The hidden units of a BM mediate interactions between the visible units and serve as internal representations of the data. 

A successfully trained BM can capture salient features of the input data. For example, the BM learns about pen strokes from an image dataset of handwritten digits~\footnote{\url{http://deeplearning.net/tutorial/rbm.html}}. Once trained, the BM can generate new samples from the learned distribution. 
%Since the simulation of the BM is cheaper than the original physical models one can accumulate the update to decorrelated the samples. Moreover, BM can discover something fundamentally new. The advantage of using BM is that by learning the salient features into the BM, can generate those updates more intelligently. Since the BM discovers more efficient updates.
%The basic idea of Ref.~\cite{Huang:2017fg} is that to train the BM to learn the probability distribution of statistical physics model.
%Given the parameters of the BM, one can of course perform Monte Carlo sampling of it to obtain samples of the visible units. Sampling according to \Eq{eq:acc} will produce statistically exact results because it respects the detailed balance condition of the physical distributions~\cite{SM}.
%The acceptance rate \Eq{eq:acc} is legitimate in general for any parameters of the BM.  
%Generative
To keep the physical simulation unbiased, the BM recommended update of the visible units $\mathbf{s}\rightarrow \mathbf{s}'$ %$T(\mathbf{s} \rightarrow\mathbf{s}')=F_\mathbf{h}(\mathbf{s}\rightarrow \mathbf{s}') p(\mathbf{h}|\mathbf{s})$ 
is accepted with the probability according to Metropolis-Hastings rule~\cite{Metropolis:1953in, Hastings:1970aa, SM}, 
\begin{equation}
%A (\mathbf{s}\rightarrow \mathbf{s}')  = \min\left \{1,  \prod_{\ell}\left(\frac{1+e^{Wx_{\ell} +b}}{1+e^{Wx'_{\ell}+b}}\cdot\frac{e^{Kx'_{\ell} }}{e^{Kx_{\ell} }}\right)  \right\}
A (\mathbf{s}\rightarrow \mathbf{s}')  = \min\left[\,1, \; \frac{p(\mathbf{s})}{p(\mathbf{s}')}\cdot\frac{\pi(\mathbf{s}')}{\pi(\mathbf{s})}\, \right]. 
\label{eq:acc}
\end{equation}
%where $\pi(\mathbf{s})$ is the unnormalized statistical weight of the physical model. 
Equation (\ref{eq:acc}) shows that the BM guides the simulation and increases the acceptance rate by exploiting the knowledge learned from data. In particular, one can even achieve a rejection free Monte Carlo simulation scheme if the BM perfectly describes the target probability distribution. This is in principle possible because BM is a universal approximator of discrete probability distributions~\cite{Freund:1994tu, LeRoux:2008ex, Montufar}. %However, extending its expressibility by including a larger number of hidden variables and more complex connections also increases the complexity of learning and sampling. 
The expressive powers of BM were studied recently from physics perspectives~\cite{Chen:2017ta, Deng:2017um, Gao:2017uk, Huang:2017ud}. See Refs.~\cite{Carleo:2016vp, Torlai:2016bm, Anonymous:aB8JAtFF} for other recent applications of BM to quantum and statistical physics problems. 

%Why RBM is not sufficient 
Reference \cite{Huang:2017fg} argues that the efficient simulation of the BM, and more importantly, its ability to capture high level features make BM ideal recommender systems to accelerate Monte Carlo simulation of challenging physical problems. In particular, Ref.~\cite{Huang:2017fg} employs a restricted architecture of BM where the connections are limited to be between the visible and hidden units. Such a restricted BM can be sampled efficiently by blocked Gibbs sampling alternating between the hidden and visible units. The generative sampling of the restricted BM already gives rise to nonlocal updates because the BM learns about collective density correlations from the Monte Carlo data. This suggests that besides being used as a general purpose recommender engine for accelerating Monte Carlo simulations, the BM may discover conceptually new efficient updates.
%which besides being a general purpose recommender engine for Monte Carlo simulations of physical systems. 
%Since the recommender updates are generated via Monte Carlo simulation of the BM themselves, we discuss this first in the following. We have just described a general and efficient approach to simulate the 3-way RBM \Eq{eq:3wayRBM}. Using it we can  recommend the RBM update to the simulation of physical models~\cite{Huang:2017fg}.
%To demonstrate power of the 3-way RBM, we 
%Consider the following rule to make transition for the Ising configurations.
%We initialize the visible spins of the 3-way RBM with the current Ising configurations, then perform one step of update according to Fig.~\ref{fig:3wayRBM}(b). 
%Our final goal is, however, not to simulate this artificial construction but to use the simulation of the 3-way RBM to
% In particular we illustrated that the BM can discover efficient cluster update using the Ising model and its generalization.
%outline of this paper
%Architecture is important

In this paper, we demonstrate the BM's creative power by exact constructions and then present a general framework exploit the power. The crucial insight is that the hidden units of the BM can learn to mediate complex interactions between the visible units and decouple the visible units into disconnected clusters. The generative sampling of the BM then automatically propose efficient cluster updates. 
%Sampling from these general BM can still be done efficiently. 
To encourage these discoveries, it is crucial to design the BM in a suitable architecture and allow its parameters adapt to the physical distribution via learning.  
%we show that besides being a general purpose recommender engine for Monte Carlo simulations of physical systems, a BM can also discover efficient cluster Monte Carlo algorithms.

%Ref.~\cite{Liu:2017da} made similar attempts using classical spin models and molecular gases. 
%There questions has been that with simpler reference has very limited expressive power (except for the simple demo studied in Refs....), while for a more sophisticated model such as a dense RBM with dense connections it it any probability description it is harder to train and sample from.

%Compared to the block Gibbs sampling of the RBM, here the three spin interaction renders the conditional probability $p(\mathbf{s}|\mathbf{h})$ intractable in general. 

%In this paper first show that the BM \emph{can} discover efficient cluster update using the Ising model and its generalization as examples. Finally, we present a general framework to train BM with MC data. 

%The paper present further developments build on Ref.~\cite{Huang:2017fg}. We design a simpler yet more powerful BM architecture with much fewer parameters and respect the translational invariance of the physical model. More importantly, sampling of the RBM automatically restores the cluster updates~\cite{1987PhRvL..58...86S} for general physical systems it was trained for. Generalization of the setup offers a general design pattern for cluster updates. 

\paragraph{Example: Ising model--}
To make the discussions concrete, we start with the classical Ising model and show that the generative sampling of the BM encompasses a wide range of celebrated cluster algorithms~\cite{1987PhRvL..58...86S, Niedermayer:1988eo, Kandel:1991dy, Kawashima:1995je, Higdon:1998fz}. The Boltzmann weight of the Ising model reads
\begin{equation}
\pi(\mathbf{s})=\exp \left( \beta J \sum_{\ell}\prod_{i \in \ell} s_{i} \right), 
\label{eq:Ising}
\end{equation}
where $\beta=1/T$ is the inverse temperature and $J$ is the coupling constant. We consider ferromagnetic coupling $J>0$ in the following for clarity. The considerations are nevertheless general and valid for the antiferromagnetic case as well. Equation (\ref{eq:Ising}) consists of a summation over links $\ell$ of a lattice and a product over Ising spins $s_{i}\in \{-1, 1\}$ reside on the vertices connected by the link. 
%There are known efficient cluster Monte Carlo updates~\cite{xxx} of the Ising model. 
%/Z_\mathrm{Ising}$, where $Z_\mathrm{Ising}=\sum_{\mathbf{s}}w(\mathbf{s})$ is the partition function of the Ising model. 

\begin{figure}[!t]
\centering
\includegraphics[width=\columnwidth]{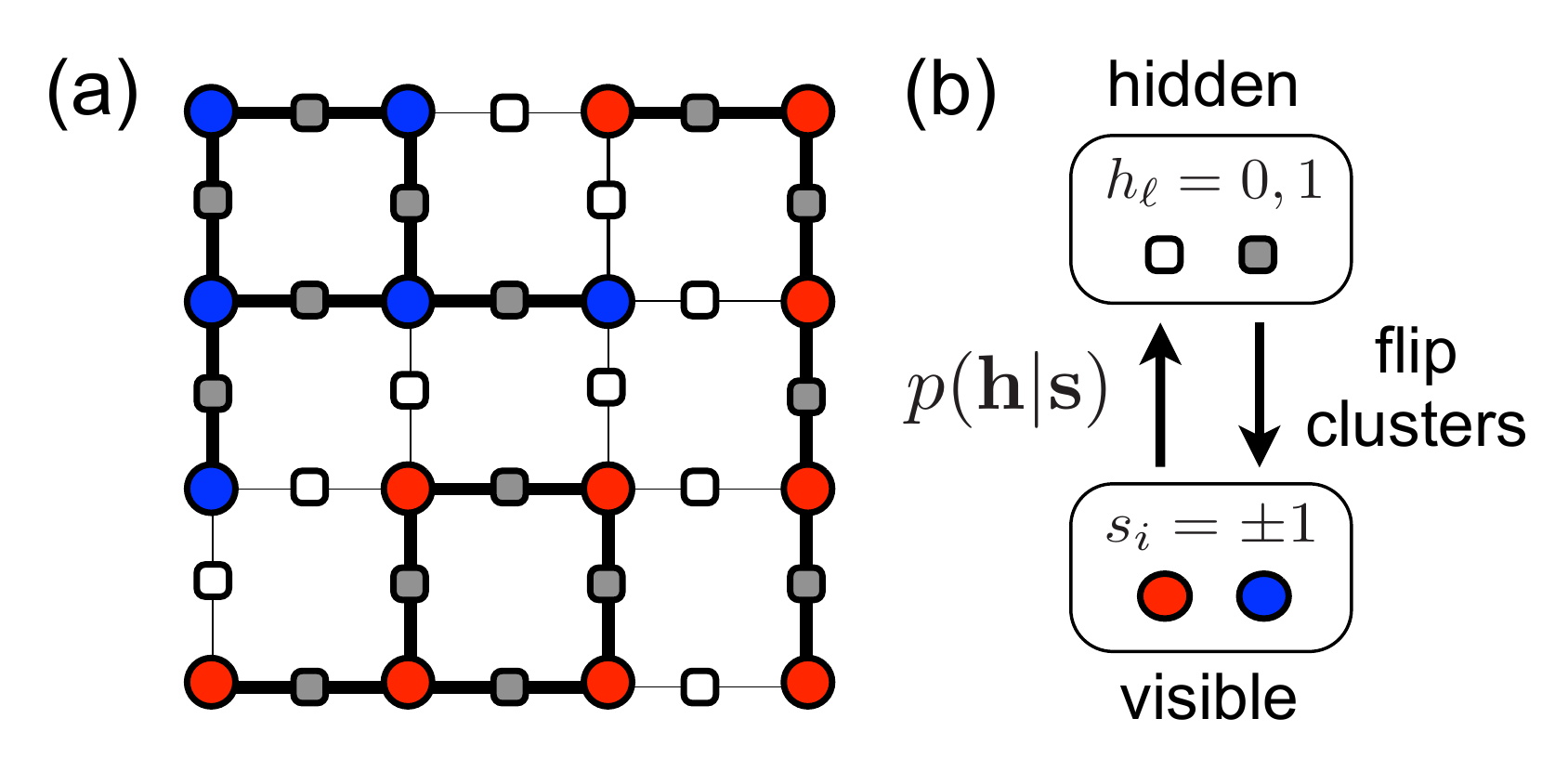}
\caption{(a) The Boltzmann Machine \Eq{eq:3wayRBM} reproduces cluster Monte Carlo algorithms of the Ising model \Eq{eq:Ising}. Solid dots  reside on the vertices are the visible units representing the Ising spins. Red and blue colors denote Ising spin up and down. The squares reside on the links are the binary hidden units, where white and gray color indicates inactive ($h_{\ell}=0$)  or active ($h_{\ell}=1$) status of the hidden unit. The effective interaction between the visible units can  either be $W$ (thick links) or $0$ (thin links). (b) The sampling of the BM. Given the visible units, we sample the hidden units according to \Eq{eq:s2h}. The inactive hidden units (white squares) divide the visible units into disconnected components which can be flipped collectively at random.}
\label{fig:3wayRBM}
\end{figure} 

To devise a BM inspired cluster update of the Ising model, we consider the architecture illustrated in Fig.~\ref{fig:3wayRBM}(a). We view the Ising spins as visible variables and introduce binary hidden variables $h_{\ell}\in \{0, 1\}$ on the links of the lattice. These units are coupled according to the following energy function 
\begin{equation}
%E (\mathbf{s}, \mathbf{h})  = -W\sum_{\ell}  h_{\ell} s_{i_{\ell}} s_{j_{\ell}}  -b\sum_{\ell}  h_{\ell} 
E(\mathbf{s}, \mathbf{h})  = -\sum_{\ell} \left(W \prod_{i\in \ell}s_{i} +b \right) h_{\ell}.  
%Z = \sum_{\mathbf{s},\mathbf{h}}\prod_{\ell} e^{\left(Ws_{i_{\ell}} s_{j_{\ell}} +b \right) h_{\ell}},   
\label{eq:3wayRBM}
\end{equation}
Equation (\ref{eq:3wayRBM}) is a high-order BM~\cite{Sejnowski:1986tm} because the interaction consists of three-spin interactions (one hidden unit and two visible units). Similar architectures were discussed in the machine learning literature under the name three-way Boltzmann Machines~\cite{Memisevic:2007vg, Memisevic:2010go, Krizhevsky:2010va}. In light of the translational invariance of the Ising model (\ref{eq:Ising}) we use the same connection weight $W$ and bias $b$ for all the links. Therefore the BM energy function \Eq{eq:3wayRBM} only contains two free parameters. 

%The summation is over the edge of the graph, and the visible spins live on the vertex, while the hidden units live on the bond of the graph. 
To perform generative sampling of the BM (\ref{eq:3wayRBM}) we proceed in two steps by exploiting its particular architecture shown in Fig.~\ref{fig:3wayRBM}. First, given a set of visible Ising spins, we can readily perform direct sampling of the hidden units. This is because the conditional probability factorizes into products over each link $p(\mathbf{h}|\mathbf{s}) = p(\mathbf{s},\mathbf{h})/p(\mathbf{s}) = \prod_{\ell} p(h_{\ell}| \mathbf{s})$, where 
\begin{equation}
 p(h_{\ell}=1| \mathbf{s}) = \sigma\left(W\prod_{i\in \ell }s_{i} +b\right),%=\frac{1}{1+\exp{\left[-\left(W\prod_{i\in\ell}s_{i} +b\right)\right]}}
 \label{eq:s2h}
\end{equation}
and $\sigma(z)=1/(1+e^{-z})$ is the sigmoid activation function. %Physically, it means once the physical spins are set, one can sample each hidden unit independently with \Eq{eq:s2h}. 
%On the other hand, different from the simpler RBM with only two-body interactions between the visible and hidden units~\cite{Huang}, one can not easily sampling back from the hidden to the visible units because of the 3-spin interaction. 
As shown in Fig.~\ref{fig:3wayRBM}(a) the inactive hidden units (white squares) divide the lattice into disconnected components since $h_{\ell}=0$ in \Eq{eq:3wayRBM} decouples the visible Ising spins reside on the link $\ell$. One can therefore identify connected components using union-find algorithm~\cite{sedgewick2011algorithms, gubernatis2016quantum} and flip all the visible Ising spins within each component collectively at random. This cluster move respect the statistical weight of the BM \Eq{eq:3wayRBM} due to the $Z_{2}$ symmetry of visible Ising spins in the energy function. 

Combining the two steps in Fig.~\ref{fig:3wayRBM}(b) forms an update of the visible units of the BM. Recommending the update to the Ising model Monte Carlo simulation, it is accepted with the probability \Eq{eq:acc}~\cite{SM}. Matching the unnormalized marginal distribution of the BM (\ref{eq:3wayRBM})
$
p(\mathbf{s}) = \prod_{\ell}\left (1+e^{W\prod_{i\in \ell}s_{i} +b}\right)
$
%where we introduced a bond variable $x_{\ell}= s_{i_{\ell}} s_{j_{\ell}}$ for the spin product. These bond variables are of course not independent. We use them only as a short hand notations in this paper. 
%By choosing the RBM architecture with links defined for the nearest neighbor sites and and tuning its parameters, one can optimize the acceptance rate when the distributions. For this we need to train the RBM  to let it learn about the physical model.  
and the Ising model  Boltzmann weight \Eq{eq:Ising} gives rise to a rejection free Monte Carlo scheme. The resulting condition  %, where $K=\beta J $ of the Ising model. 
\begin{equation}
\frac{1+e^{b+W}}{1+e^{b-W}} = e^{2\beta J}
\label{eq:IsingWb}
\end{equation}
can always be satisfied with appropriate chosen $W$ and $b$. 
%With two unknowns in the equation there is always a solution. The solutions of \Eq{eq:IsingWb} read $W +b =  \ln\left[(r\pm\sqrt{r^{2} + 4(r+1)e^{2b}}) /2\right] $, where $r= e^{2\beta J}-1$. 
%If we require the \Eq{eq:IsingWb} to be satisfied, the bias on the hidden unit $b$ becomes the only parameter of the algorithm. %Figure~\ref{fig:Wb} shows the solutions for different ferromagnetic Ising couplings $K>0$. 
It is instructive to examine the BM recommended updates in two limiting cases. 
 
% The Gibbs sampling becomes to a collective cluster flip.
%\begin{figure}[!t]
%\centering
%\includegraphics[width=\columnwidth]{Wb.pdf}
%\caption{The solution of \Eq{eq:IsingWb} for various Ising coupling strengths.}
%\label{fig:Wb}
%\end{figure} 

In the limit of $b\rightarrow -\infty$, the solution of \Eq{eq:IsingWb} reads $W+b = \ln(e^{2 \beta J}-1)$. Thus, the conditional sampling of the hidden units \Eq{eq:s2h} will set $h_{\ell}=1$ with probability $\sigma(W+b)=1-e^{-2\beta J}$ if the link connects to two parallel spins $\prod_{i\in\ell}s_{i}=1$. While it  will always set the hidden unit to inactive $h_{\ell}=0$ if the link connects to anti-parallel spins. 
%As illustrated in Fig.~\ref{fig:3wayRBM}(a), the links with inactive hidden units $h_{\ell}=0$ breaks the system into isolated regions which can be flipped freely. 
Combined with the random cluster flip of visible units, this BM recommended update shown in Fig.~\ref{fig:3wayRBM}(b) exactly reproduces the Swendsen-Wang cluster algorithm~\cite{1987PhRvL..58...86S} of the Ising model.  

While in the opposite limit $b\rightarrow \infty$, the solution of \Eq{eq:IsingWb} approaches to $W=\beta J$. In this limit, all the hidden units are frozen to $h_{\ell}=1$ because the activation function in \Eq{eq:s2h} saturates no matter whether the visible Ising spins are aligned or not. The BM \Eq{eq:3wayRBM} then trivially reproduces the Ising model statistics by copying its coupling constant $\beta J$ to the connection weight $W$. In this limit the BM recommended update shown in Fig.~\ref{fig:3wayRBM}(b) is a trivial global flip of the visible Ising spins.  
%\begin{figure}[!t]
%\centering
%\includegraphics[width=\columnwidth]{Ising.pdf}
%\caption{Results obtained for a two dimensional Ising model on a $10\times 10$ square lattice at the critical point $\beta=\beta_{c}$. (a) The acceptance rates reaches one when the \Eq{eq:IsingWb} is satisfied. (b) The averaged cluster size measure in the fraction of the lattice size. (c) The energy autocorrelation time improvs by a fact of $10^{4}$ with optimal choice of the weight and bias of the 3-way RBM.}
%\label{fig:Ising}
%\end{figure} 

In between the above two limiting cases, the BM still recommends valid rejection free Monte Carlo updates for the Ising model. These updates correspond to the Niedermayer's cluster algorithm~\cite{Niedermayer:1988eo} where the sites are randomly connected into clusters according to \Eq{eq:s2h} and the clusters may contain misalligned visible spins. The bias parameter $b$ in \Eq{eq:s2h} controls the activation threshold of the hidden units and thus affects the average cluster size. In essence, the BM (\ref{eq:3wayRBM}) encompasses several general 
cluster Monte Carlo frameworks including the Kandel-Domany~\cite{Kandel:1991dy} and the dual Monte Carlo~\cite{Kawashima:1995je, Higdon:1998fz} algorithms. In these algorithms, the Monte Carlo sampling alternates between the physical degrees of freedom and auxiliary graphical variables corresponding to the hidden units of the BM. %However, the simple model \Eq{eq:3wayRBM} covers a large number of ingenious cluster algorithms~\cite{1987PhRvL..58...86S, Niedermayer:1988eo, Kandel:1991dy, Kawashima:1995je, Higdon:1998fz}.

%The above discussions show that the BM with suitable chosen architecture and parameters can reproduce the cluster updates~\cite{1987PhRvL..58...86S, Niedermayer:1988eo, Kandel:1991dy, Kawashima:1995je, Higdon:1998fz}. 

%However, it should be emphasized that these parameters are learnable, in the sense one can obtain their values with a data driven approach.
%However, even this extremely simple RBM capture important collective correlations of the model. See, Fig~\ref{fig:fit} show the fitting is also reasonably good.  

%Figure~\ref{fig:Ising} shows the results obtain for the recommended update as a function of the weight $W$ for various biases $b$. The acceptance rate reaches one when the \Eq{eq:IsingWb} is satisfied. At the same point the autocorrelation time shows a dip, indicating the optimal performance at a given bias value.  The overall performance can differ by a factor of $10^{4}$ with suitable choice tuning of the RBM parameters. Positive bias promote the hidden unit to much so there are too few clusters. The autocorrelation time measured in the unit of sweeps~\cite{NewmanBook}, and computed using the efficient algorithms of \cite{Ambegaokar:2010kj}. The average cluster size decreases for more negative bias since this discourages the visible variables to connect into components. The example shows that using finite negative (no need to be minus infinity) bias already achieve quite good performance. And that the requirement need not to be satisfied perfectly. 

%JK Model%%%%%%%%%%%%%%%%%%%%%%% 
\paragraph{Example: Ising model with plaquette interactions--}
The potential of BM goes beyond reproducing existing algorithmic frameworks~\cite{1987PhRvL..58...86S, Niedermayer:1988eo, Kandel:1991dy, Kawashima:1995je, Higdon:1998fz}. By further exploiting its power from latent representations one can make nontrivial algorithmic discoveries. We illustrate this using the plaquette Ising model~\cite{Liu:2017da} as an example. The Boltzmann weight reads
\begin{equation}
\pi (\mathbf{s})= \exp{ \left( \beta J \sum_{\ell} { \prod_{i\in \ell}s_{i}  } + \beta K \sum_{\wp}{  \prod_{i\in \wp} s_{i} } \right)}, 
\label{eq:JKmodel}
\end{equation}
where the second term contains four-spin interactions on each square plaquette denoted by $\wp$. We consider $K>0$ for concreteness. Since no simple and efficient cluster algorithm is known, Ref.~\cite{Liu:2017da} fits the Boltzmann weight \Eq{eq:JKmodel} to an ordinary Ising model \Eq{eq:Ising} and propose Monte Carlo updates by simulating the latter model with cluster algorithms~\cite{Wolff:1989iy, PhysRevLett.71.2070}. However, the acceptance rates decrease for large systems due to imperfect fittings and the approach ends up to show similar scaling behavior as the single spin flip update. 

\begin{figure}[!t]
\centering
\includegraphics[width=\columnwidth]{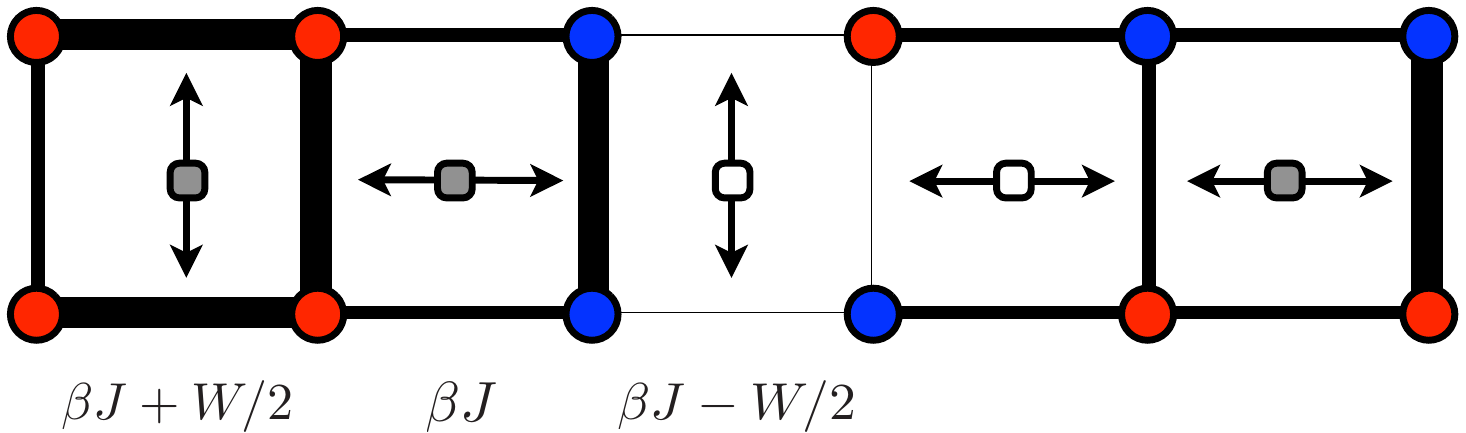}
\caption{The Boltzmann Machine \Eq{eq:BMsh} suggests a new cluster update for the plaquette Ising model (\ref{eq:JKmodel}). Red/blue dots on the vertices denote the visible Ising spins, and white/gray squares in the plaquette center denote the hidden units. The double arrows indicates two parallel links $\ell_{\wp}, \bar{\ell}_{\wp}$ composing of the plaquette $\wp$. The hidden units are sampled directly according to \Eq{eq:s2hp} where the break up of the plaquette into parallel links are chosen at random. Once the hidden units are given, \Eq{eq:BMsh} reduced to an inhomogeneous Ising model where the visible spins are coupled with modified coupling strengths indicated by the thickness of the links.}
\label{fig:plaq}
\end{figure} 

Here we construct a BM which suggests an efficient, unbiased, and rejection free cluster Monte Carlo algorithm for \Eq{eq:JKmodel}. First, we decompose the four-spin plaquette interaction using the Hubbard-Stratonovich (HS) transformation~\cite{stratonovich1957method, Hubbard:1959vo} 
\begin{equation}
 \exp\left( \beta K \prod _{i\in \wp}s_{i} \right)  = \frac{e^{-\beta K}}{2}\sum_{h_{\wp}\in \{0, 1\}}\exp \left[W \left(h_{\wp}-\frac{1}{2}\right)\mathcal{F}_{\wp}(\mathbf{s})\right], 
 \label{eq:HS}
\end{equation}
%\begin{eqnarray}
% \exp\left( \beta K \prod _{i\in \wp}s_{i} \right) & = &  \exp\left( \beta K \prod _{i\in \ell_{\wp}}s_{i} \cdot\prod _{i\in \bar{\ell}_{\wp}}s_{i} \right)  \label{eq:HS} \\ &= &  \frac{e^{-\beta K}}{2}\sum_{h_{\wp}\in \{0, 1\}}\exp \left[W \left(h_{\wp}-\frac{1}{2}\right)\mathcal{F}_{\wp}(\mathbf{s})\right], \nonumber 
%\end{eqnarray}
where $ W = \mathrm{acosh}(e^{2\beta K })$ is the coupling strength between the binary HS field $h_{\wp}$ and the sum of two-spin products $\mathcal{F}_{\wp}(\mathbf{s}) =  \prod_{i\in \ell_{\wp} } s_{i} + \prod_{i\in \bar{\ell}_{\wp} } s_{i}$ defined for the plaquette. The two parallel links $\ell_{\wp}$ and $ \bar{\ell}_{\wp}$ constitute the plaquette $\wp$, see Fig.~\ref{fig:plaq}. %because $\mathcal{F}_{\wp}(\mathbf{s})= \mathcal{F}^{3}_{\wp}(\mathbf{s})$ and $\mathcal{F}^{2}_{\wp}(\mathbf{s})= \mathcal{F}^{4}_{\wp}(\mathbf{s})= (1+\prod _{i\in \wp}s_{i} )/2$.
Equation (\ref{eq:HS}) is equivalent to the discrete HS transformation widely adopted for the Hubbard models~\cite{1983PhRvB..28.4059H}. Regarding the HS field $h_{\wp}$ as a hidden unit, the following BM
\begin{equation}
E(\mathbf{s}, \mathbf{h}) = -\sum_{\ell} \left[\beta J + W\sum_{\wp} \left(h_{\wp}-\frac{1}{2}\right)\left(\delta_{\ell \ell_{\wp}} + \delta_{\ell \bar{\ell}_{\wp}} \right)\right]\prod_{i\in\ell} s_{i}
\label{eq:BMsh}
\end{equation}
exactly reproduces \Eq{eq:JKmodel} after marginalization. %Figure~\ref{fig:plaq} illustrates structure of  the BM. 
Since \Eq{eq:HS} holds for arbitrary partition of the plaquette into two links $\ell_{\wp} \cup \bar{\ell}_{\wp}=\wp$ and $ \ell_{\wp} \cap \bar{\ell}_{\wp} = \varnothing$, we choose vertical or horizontal break up at random for each plaquette. 
 
Simulation of the BM \Eq{eq:BMsh} suggests an efficient cluster update for the original plaquette Ising model (\ref{eq:JKmodel}). First of all, sampling the hidden variables given the visible Ising spins is straightforward since the conditional probability factorizes over plaquettes $p(\mathbf{h}|\mathbf{s}) = \prod_{\wp} p(h_{\wp}|\mathbf{s})$, where 
\begin{equation}
p(h_{\wp}=1|\mathbf{s}) = \sigma \left(W\mathcal{F}_{\wp}(\mathbf{s}) \right). 
\label{eq:s2hp}
\end{equation}
Therefore, the hidden unit of each plaquette activates independently given the local feature $\mathcal{F}_{\wp}(\mathbf{s})$. Next, once the hidden variables are given, the BM \Eq{eq:BMsh} corresponds to an Ising model with two-spin interactions only, shown in Fig.~\ref{fig:plaq}. One can sample it efficiently using the cluster updates~\cite{1987PhRvL..58...86S} by taking into account of the randomly modified coupling strengths. As discussed in the above, this amounts to introduce another set of hidden units which plays the role of auxiliary graphical variables. %In total, the cluster update are learned with two sets of hidden variables. 
Finally, according to \Eq{eq:acc} the updates of the visible Ising spins are always accepted because the BM \Eq{eq:BMsh} exactly reproduces the statistics of the plaquette Ising model \Eq{eq:JKmodel}.

\begin{figure}[!t]
\centering
\includegraphics[width=\columnwidth]{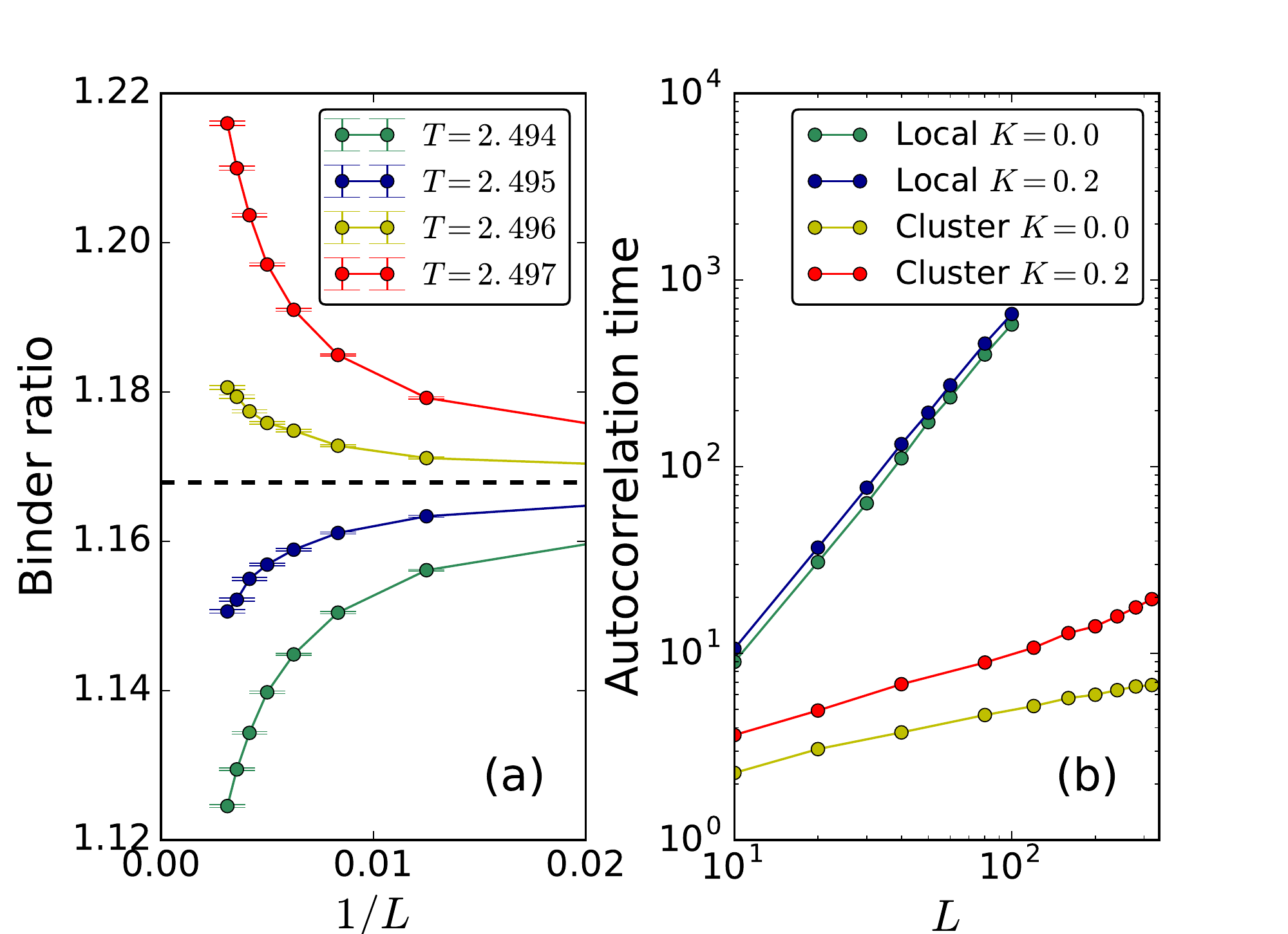}
\caption{Results for the Ising model with four-spin plaquette interactions  (\ref{eq:JKmodel}) on square lattices with linear length $L$. (a) Binder ratio obtained using the cluster update suggested by BM \Eq{eq:BMsh} at $K/J=0.2$. The dashed line indicates the universal value for the two-dimensional Ising universality class. (b) The cluster update improves the energy autocorrelation time by orders of magnitude at the critical point compared to the local update. }
\label{fig:JK}
\end{figure}

To demonstrate the efficiency of the discovered cluster update, we simulate the plaquette Ising model (\ref{eq:JKmodel}) in the vicinity of the critical point and compare performance to the simple local update algorithm. Figure~\ref{fig:JK}(a) shows the Binder ratio $\braket{\left(\sum_{i}s_{i}\right)^{4}}/\braket{\left(\sum_{i}s_{i}\right)^{2} }^{2}$ for various system sizes at $K/J=0.2$, which indicates a critical temperature $T/J=2.4955(5)$. The black dashed line indicates the universal critical value of the Binder ratio $1.1679$ corresponding to the two-dimensional Ising universality class~\cite{Salas:1999qh}.
%The average cluster size approaches to a constant value independent of the system size. 
Figure~\ref{fig:JK}(b) shows the energy autocorrelation times~\cite{Ambegaokar:2010kj} of the local updates and the cluster updates at the critical point, both measured in the unit of Monte Carlo sweeps of the visible spins~\cite{Newman:1999fd}. The local updates exhibit the same scaling for the Ising model ($K=0$) and the plaquette Ising model ($K=0.2$). While the cluster updates are orders of magnitude more efficient than the local updates. The dynamic exponent of the cluster algorithm is also significantly reduced compared to the local update. %In future, it is interesting to apply the discovered cluster Monte Carlo algorithm to Ising gauge field theory~\cite{RevModPhys.51.659}.

%We identify $s_{i}$ as the visible units of the 3-way RBM, the simulation is identical to the Ising model except we replace $p_\mathrm{Ising}(\mathbf{s})$ in \Eq{eq:acc} by the probability distribution the Falicov-Kimball model~ \cite{Huang:2017fg}. While we need to train the 3-way RBM to let its $p(\mathbf{s})$ approximate it. 
%Note that in \Eq{eq:3wayRBM} we have for simplicity only put the hidden units on the nearest neighbor bonds. This reproduces the known cluster update.

\paragraph{General framework--}
To sum up, we outline a general framework of discovering cluster updates using the following BM
\begin{equation}
E(\mathbf{s}, \mathbf{h}) = - E({\mathbf{s}}) -\sum_{\alpha}\left[ W_{\alpha} \mathcal{F}_{\alpha}(\mathbf{s}) + b_{\alpha} \right] h_{\alpha},
\label{eq:generalRBM}
\end{equation}
where $\mathcal{F}_{\alpha}(\mathbf{s})$ is a feature of the visible units
and $h_{\alpha}\in\{0,1\}$ is the corresponding hidden variable. $W_{\alpha}$ and $b_{\alpha}$ are connection weight and bias, and $\alpha$ is the index for various features. For example, \Eq{eq:3wayRBM} and \Eq{eq:BMsh} used the features defined on links ($\alpha=\ell$) and on plaquettes  ($\alpha=\wp$)  respectively.  
%The corresponding features are spin product on links $\mathcal{F}_{\ell}(\mathbf{s})=\prod_{i\in \ell} s_{i}$ and on plaquettes $\mathcal{F}_{\wp}(\mathbf{s})=\prod_{i\in \wp} s_{i}$. 
%On the other hand, there were cases to break \Eq{eq:fitting} to have smaller clusters. In this case the cluster flip is not rejection free anymore. 
%In general, the feature $\mathcal{F}_{\alpha}(\mathbf{s})$ can consist of visible spins at long distance % These connections can be arbitrary and independent of the original problem. %In the FK model case we consider the connections between the nearest and the next-neartest neighbors sites. There are in total $5$ fitting parameters 
%or even the multispin interactions. 
%As long as the constructed energy potential has the Z2 symmetry for the visible spins.  
In general, one is free to design features consist of long-range interactions or even multispins interactions~\footnote{Introducing two set of hidden units coupled respectively to the feature $\mathcal{F}_{\ell}(\mathbf{s})=\prod_{i\in \ell}s_{i}$ and $\mathcal{F}_{\wp}(\mathbf{s})=\prod_{i\in \wp}s_{i}$ gives an alternative cluster algorithms for the plaquette Ising model \Eq{eq:JKmodel}. The resulting algorithm is a simple generalization of the Swendsen-Wang algorithm~\cite{1987PhRvL..58...86S} to the case of plaquette unit. The satisfied plaquette $\mathcal{F}_{\wp}(\mathbf{s})=1$ is connected with probability $1-e^{-2\beta K}$ while the unsatisfied plaquette is always disconnected. However, this algorithm has larger averaged clusters size compared to the one presented in the texts. This is because the visible units are more likely to be connected with the plaquette decomposition. }. There are several crucial points in the design of \Eq{eq:generalRBM}. First, one can easily sample the hidden units conditioned on these features since there is no interaction between the hidden variables, i.e. \Eq{eq:generalRBM} is a semi-restricted BM. The activation probability of each hidden unit is $ p(h_{\alpha}=1| \mathbf{s}) = \sigma\left(W_{\alpha}\mathcal{F}_{\alpha}(\mathbf{s})  +b_{\alpha}\right)$. Second, once the hidden units are given, \Eq{eq:generalRBM} reduces to an effective model for the visible spins, which should be easier to sample compared to the original problem. For example, one can randomly flip each disconnected component separated by the inactive hidden units if $E(\mathbf{s})=0$ and $\mathcal{F}_{\alpha}(\mathbf{s}) = \mathcal{F}_{\alpha}(-\mathbf{s})$. Alternatively, one can build another BM to simplify the sampling of \Eq{eq:generalRBM} given the hidden units. 
One can even apply this idea iteratively and build a hierarchy of BMs. Overall, the key design principle is to choose appropriate features in \Eq{eq:generalRBM} such that the BM correctly reproduce the distribution of the physical problem and is easy to simulate. A good design is likely to exploit the knowledge of the original physical problem \footnote{When all the hidden units are frozen to the active state by infinitely strong bias, the marginal probability of the BM (\ref{eq:generalRBM}) approaches to $ p(\mathbf{s}) = e^{ E(\mathbf{s})+ \sum_{\alpha}W_{\alpha} \mathcal{F}_{\alpha}(\mathbf{s}) } $. In this limit the BM reduces to an effective Hamiltonian of the visible spins with designed features.  
One thus loses the flexibility empowered by the latent structures. Moreover, lacking the nonlinearity in the marginal distribution induced by the dynamical hidden units also reduces the fitting capacity of the BM.}. 

%The general feature of these BMs are that one can sample the hidden units directly given the visible variables, as their conditional distribution only depends on local units. While given the hidden variables, the BM is a simple model, for example the visible untis breaks into domains which can flip freely, or the BM is a where there is known efficient updates.
 %and require all parameters are equal we restore the simple 3-way RBM \Eq{eq:3wayRBM} used in the actual calculation. 

%To match the marginal probability distribution to the one of plaquette Ising model \Eq{eq:JKmodel} lead to equations similar to \Eq{eq:IsingWb}. 

%In this case, one can perform multiple steps of the RBM sampling before suggests it back to the physical simulation. Since the simulation of the RBM is much cheaper than the simulation, 

%\begin{figure}[!t]
%\centering
%\includegraphics[width=\columnwidth]{FK.pdf}
%\caption{Results for the Falicov-Kimball model  (\ref{eq:FKmodel}). (a) The acceptance rate (b) The Binder ratio. The dashed line indicates the universal value for the 2D Ising universality class. (c) The autocorrelation time of the Binder ratio.}
%\label{fig:FK}
%\end{figure} 

The BM \Eq{eq:BMsh} provides a general paradigm to discover efficient cluster updates automatically from data because its parameters are \emph{learnable}, see Fig.~\ref{fig:concept}. 
Although we adopted a constructive approach in this paper, in general, the hidden units of the BMs can learn to be auxiliary graphical variables or HS fields. The BM learning can be done in multiple ways, either through unsupervised learning of the configuration data~\cite{Hinton:2002ic, Torlai:2016bm}, or through supervised learning of the unnormalized target distribution $\pi(\mathbf{s})$~\cite{Huang:2017fg}, or even through reinforcement learning~\cite{Sutton:1998} by optimizing the autocorrelation time of the Monte Carlo samples. 
 %Although the training may not be perfect in the absence of the analytical solutions like \Eq{eq:IsingWb} or \Eq{eq:HS}, the physical results are unbiased as long as one accepts these recommended updates according to \Eq{eq:acc}. 

%One additional feature we did not touch on too much are that the connection weight and bias of the BM are learnable. 
% To determine the parameters of the generalized RBM, one can perform the fitting of the log-probability of the physical model as was done in~\cite{Huang:2017fg}. 
%Therefore, one should encourage negative bias in the training so that the lattice breaks into smaller clusters. 
%In this case the hidden units play a dual role of HS field and the one to break the lattice. 
%There are three ingredients of BM, architecture, learning and sampling. In this paper we focus on the first two, because the learning is done exactly and analytically for the simple system at consideration. 
%In the sense given the sample data one can performs learning algorithms (either supervised or unsupervised) to learn the weight. 
%Distill into human knowledge and understanding. 
In closing, we note many cluster quantum Monte Carlo algorithms~\cite{Evertz:2003ch, Kawashima:2004clb} share the framework of~\cite{Kandel:1991dy, Kawashima:1995je, Higdon:1998fz}. Generalization of the hidden units to higher integers or even continuous variables is likely to increase the capacity of the BM.  One can include higher order self-interactions of the hidden variables in \Eq{eq:generalRBM} in this case. To this end, these BMs provide concrete parametrization of valid Monte Carlo update policies which can be optimized through learning. This approach opens a promise of discovering practically useful Monte Carlo algorithms for a broad range of problems, such as frustrated magnets or correlated fermions where known efficient cluster updates are rare. Exploring more general and powerful BM architectures in these settings may lead to even more exciting algorithmic discoveries.

%Before arriving at \Eq{eq:generalRBM} may be introduced other hidden units to decouple the interactions such as done in \Eq{eq:HS}. In this case the parameters $W, b$ also depends on the value of the HS fields. 

%Linking to the known cluster update literature demonstrated that the present approach is powerful enough which can already reproduce the most significant cluster updates. Therefore, understanding the role of hidden units shows that one can see BM not only accelerate quantum Monte Carlo simulation but also offer a general framework for devising new efficient updates. Distill into human knowledge and understanding. 

%(How to numerically learn BM of BM ) ???  (May be in two steps, first learn a HS-like transformation approximately with simple interactions. Then learn the W, and b of the last layer cluster decomposition.)
%Learn $W_{hp}$ as as a CNN ? 

\begin{acknowledgments}
L.W. thanks Li Huang and Yi-feng Yang for collaborations on \cite{Huang:2017fg, Huang:2016wx} and acknowledges Jing Chen, Junwei Liu, Yang Qi, Zi-Yang Meng, and Tao Xiang for useful discussions. L.W. thanks Li Huang for comments on the manuscript. 
L.W. is supported by the Ministry of Science and Technology of China under the Grant No.2016YFA0302400 and the start-up grant of IOP-CAS. The simulation is performed at Tianhe-2 supercomputer in National Supercomputer Center in Guangzhou. We use the ALPS library~\cite{BBauer:2011tz} for the Monte Carlo data analysis. 
\end{acknowledgments}
 
\bibliographystyle{apsrev4-1}
\bibliography{RBMCluster,refs}

\clearpage
\appendix 

\section{Detailed balance condition of \Eq{eq:acc}}
The acceptance probability of the recommender update from the restricted Boltzmann Machine is derived in \cite{Huang:2017fg}. We repeat the derivation for the BM considered in the main texts for the convenience of the readers. 

First of all, the Metropolis-Hastings~\cite{Metropolis:1953in, Hastings:1970aa} acceptance rate of the physical model satisfies
\begin{equation}
A(\mathbf{s}\rightarrow \mathbf{s}') = \min \left[1, \;\frac{T(\mathbf{s}'\rightarrow \mathbf{s})}{T(\mathbf{s}\rightarrow \mathbf{s}')} \cdot\frac{\pi(\mathbf{s}')}{\pi(\mathbf{s})} \right]. 
\label{eq:MHacc}
\end{equation}
The transition probability is recommended from the simulation of the BM, where we sample alternatingly between the hidden and visible units, i.e, $T(\mathbf{s}\rightarrow \mathbf{s}')=F_\mathbf{h}(\mathbf{s}\rightarrow \mathbf{s}') p(\mathbf{h}|\mathbf{s})$. Here $F_\mathbf{h}$ denotes update of the visible units given the hidden variables $\mathbf{h}$. In Ref.~\cite{Huang:2017fg} we have used $F_\mathbf{h}(\mathbf{s}\rightarrow \mathbf{s}')=p(\mathbf{s}'|\mathbf{h})$ as the conditional probability given the hidden units of the restricted BM is simply tractable. For the BMs consider in this paper, we adopted more sophisticated update such as cluster flip transition of the visible units. The requirement on $F_\mathbf{h}$ is that it respects the joint probability distribution of the BM given the hidden units
\begin{equation}
F_\mathbf{h}(\mathbf{s} \rightarrow \mathbf{s}') p(\mathbf{s}, \mathbf{h}) = F_\mathbf{h}(\mathbf{s}' \rightarrow \mathbf{s}) p(\mathbf{s}',\mathbf{h}). 
\end{equation}

The ratio of the transition probability thus satisfy  
\begin{eqnarray}
\frac{T(\mathbf{s} \rightarrow\mathbf{s}')}{T(\mathbf{s}' \rightarrow\mathbf{s})} & = &  \frac{F_\mathbf{h}(\mathbf{s}\rightarrow \mathbf{s}') p(\mathbf{h}|\mathbf{s})}{F_\mathbf{h}(\mathbf{s}'\rightarrow \mathbf{s}) p(\mathbf{h}|\mathbf{s}')} \nonumber \\
& = &  \frac{p(\mathbf{s}', \mathbf{h}) p(\mathbf{h}|\mathbf{s})}{ p(\mathbf{s}, \mathbf{h}) p(\mathbf{h}|\mathbf{s}')}   \nonumber \\ 
& =& \frac{p(\mathbf{s}')}{p(\mathbf{s})}. 
\label{eq:recommender}
\end{eqnarray}  

Substitute \Eq{eq:recommender} into the Metropolis-Hastings acceptance probability \Eq{eq:MHacc}, we obtain \Eq{eq:acc} in the main texts. 

\end{document}